\documentclass{svjour2}                    

\smartqed  

\usepackage{graphics,graphicx,epsfig,color}
\usepackage{amssymb,amsmath,amsfonts}
\usepackage{epsf,wrapfig}
\usepackage[mathcal]{euscript}
\usepackage{mathptmx}      
 \journalname{J Stat Phys}

\begin{document}

\title{Optimal prediction and natural scene statistics in the retina \thanks{Support for this work was provided by: The Alfred P.\ Sloan Foundation, a FACCTS grant from the France Chicago Center, and a Chateaubriand Fellowship to JS.}
}

\titlerunning{Prediction in the retina}        

\author{Jared Salisbury$^{1,2}$         \and
        Stephanie E.\ Palmer$^2$
}

\institute{
              $^1$ Graduate Program in Computational Neuroscience \\
              $^2$ Department of Organismal Biology and Anatomy\\
              $\phantom{^1}$ University of Chicago\\
              $\phantom{^1}$ 1027 E 57th St., Chicago, IL 60637, USA\\
              $\phantom{^1}$ \email{sepalmer@uchicago.edu}
}

\date{Received: date / Accepted: date}

\maketitle

\begin{abstract}
Almost all neural computations involve making predictions. Whether an organism is trying to catch prey, avoid predators, or simply move through a complex environment, the data it collects through its senses can guide its actions only to the extent that it can extract from these data information about the future state of the world. An essential aspect of the problem in all these forms is that not all features of the past carry predictive power. Since there are costs associated with representing and transmitting information, a natural hypothesis is that sensory systems have developed coding strategies that are optimized to minimize these costs, keeping only a limited number of bits of information about the past and ensuring that these bits are maximally informative about the future. Another important feature of the prediction problem is that the physics of the world is diverse enough to contain a wide range of possible statistical ensembles, yet not all motion is probable. Thus, the brain might not be a generalized predictive machine; it might have evolved to specifically solve the prediction problems most common in the natural environment.  This paper reviews recent results on predictive coding and optimal predictive information in the retina and suggests approaches for quantifying prediction in response to natural motion.

\end{abstract}

\pagebreak

\section{Introduction}
\label{intro}

What are the predictable components of the input to an animal's visual system in its natural environment? While the characteristics of static images have been explored in large image repositories \cite{atick+redlich_92,dong+atick_95,olshausen+field_96,ruderman_97,bell+sejnowski_97,vanhateren+al_98,stephens+al_13}, some, but not as much is known or measured in the temporal domain \cite{dong+atick_95,billock+al_01}. One interesting feature of scaling in static images is the power-law distribution of spatial variation in local contrast \cite{dong+atick_95,ruderman_97}. This scaling implies that natural images are scale-free, displaying the same basic structure on all length scales. Power-law behavior in the frequency distribution of temporal fluctuations in total scene luminance have also been observed in a variety of natural contexts, and scenes display slightly different exponents depending on their specific content \cite{billock+al_01}. This paper will review recent attempts to connect natural motion statistics to efficient prediction in the visual system, focusing on the retina. Tying temporal statistics of natural scenes to neural prediction will reveal what types of motion the brain can efficiently represent and therefore constrain the types of predictions the brain can perform.

The concept of efficient coding for prediction in the brain has been developed in two main ways: via theories of predictive coding \cite{srinivasan+al_82,srinivasan+rao_85,rao+ballard_99,hosoya+al_05,kilner+al_07,bastos+al_12} that eliminates redundancy in the temporal response of the brain, and through analytical work to characterize the optimal trade-offs between representing the past and future sensory input  \cite{bialek+al_01,bialek+al_06} (via information bottleneck calculations\cite{tishby+al_99,chechik+al_05,creutzig+al_09}).  In this paper, we will review and relate these two approaches to neural coding in the retina, and propose methods for extending this work to the context of natural motion statistics.

It has been shown that retinal ganglion cells (RGCs), the output cells of the retina whose axons form the optic nerve, display a whole host of nonlinear processing characteristics that may be connected to prediction. RGCs respond differentially to object versus background motion \cite{olveczky+al_03}. Ganglion cells have also been shown to code for a variety of motion features in ways that cannot be accounted for by a simple receptive field picture of encoding. This includes: motion anticipation, the coding in the retina for the anticipated position of an object moving at constant velocity \cite{berry+al_99}; the omitted-stimulus response, in which ganglion cells fire after the cessation of a sequence of visual flashes at the appropriate delay where the next flash in the sequence would be expected \cite{schwartz+al_07}; and reversal responses, where neurons in the retina fire a synchronous burst of activity after the reversal of a moving bar, irrespective of their relative receptive field positions \cite{schwartz+al_07_reversal}. All of these adaptive motion-processing features speak to the retina's complexity as an encoding device, and have a component of predictability of the future state of visual stimuli. Recent work has also shown that the retina solves a general prediction problem in a near-optimal way\cite{palmer+al_15}. 

We will review this background material and discuss how extensions of this work could reveal how an organism's ecological niche shapes its predictive processing. In particular, it may be that the retinas of different species possess the capacity to solve different suites of motion prediction problems tailored to their natural environment. Evolution may shape which problems are hardwired in the early visual system, and exploring that can uncover just how far the retina is able to tune its predictive power to the statistics of its inputs.

\section{Theories of optimal prediction}
\label{optimal}
Sitting at the front end of the visual system and with a limited number of fibers to transmit all the visual information the brain receives, the retina has long been hypothesized to be an efficient and perhaps even optimal encoder of the visual world \cite{barlow_53,barlow_61,atick+redlich_92,hosoya+al_05,doi+al_12}.  This notion of efficiency dictates that the retina's representation of the visual inputs to the photoreceptor layer should be as loss-less as possible, given the number of cables the retina has along which to transmit information to the brain and the fact that metabolic constraints limit the firing rate of neurons.  To make the best use of each fiber, these signals should be independent in space and in time. Recent work from a variety of researchers expand on this simple notion of efficiency.  Natural inputs to the retina are non-Gaussian\cite{dong+atick_95,ruderman_97}, the noise spectrum in neural data is not white, retinal firing is certainly highly redundant \cite{puchalla+al_05}, and not all information about the input is equally relevant to the organism.  The concept of optimizing the predictive capacity of the retina assigns value to particular bits of information: it says that compression is only successful when the transmitted bits convey information about the future input \cite{bialek+al_06}.  The information bottleneck method \cite{tishby+al_99} is a way of defining relevant information, in this case information about the future, as the distortion measure. 

\subsection{Efficient coding in the time domain: Predictive coding}
\label{pcoding}
The efficient coding hypothesis states that all of the information about the input should be retained, while minimizing the entropy of the response \cite{shannon_48}.  If not all bits of information about the input are retained, the problem can be formulated using rate distortion theory \cite{berger_71},
\begin{equation}
\min_{p(r_t|s_t)} I(R_t;S_t) + \beta D(R_t,S_t),
\end{equation}
where $D$ is the average distortion, $R$ is the neural response to the stimulus $S$, and the minimization finds the lowest transmitted bit rate, given $D$. 


The core concept in predictive coding, in the time domain, is that temporal correlations in the output stream should be eliminated, so that only deviations from expected response, or those that are `surprising' are encoded \cite{srinivasan+al_82}.  If the input statistics are stationary, predictive coding aims to minimize the response of the system. 
The role of neurons in a predictive coding paradigm is to code for changes in response statistics, not the ongoing predictable events in a stationary world. 

Predictive coding has been postulated to be achieved through feedback connections from higher areas onto sensory input streams \cite{rao+ballard_99,kilner+al_07,wacongne+al_12,bastos+al_12}, and early \cite{srinivasan+al_82} as well as recent work in the retina hypothesizes feedforward adaptive mechanisms at the sensory periphery may result in predictive coding \cite{kastner+baccus_13}. As such, predictive coding is highly efficient, because redundancy in time is eliminated.  Mechanisms have been proposed by which the retina could implement predictive coding, via inhibitory interactions at bipolar terminals \cite{hosoya+al_05}.   Also, the work of Den\` eve shows how predictive coding may be self-organized in neural networks \cite{boerlin+al_13}.

\subsection{Predictive information}
\label{predinfo}
Information theoretic treatments of prediction in the brain focus on defining not just the code that retains the most stimulus information for a given output bit rate, but the one that retains the most information about the future stimulus.  This addition of the notion of relevant information has sharpened discussions of early sensory processing in the context of prediction \cite{bialek+al_06}.  The theory for retaining the optimal amount of predictive information has been well-developed by Tishby and colleagues \cite{tishby+al_99,chechik+al_05,creutzig+al_09}, and leads not only to elegant but also testable results.  Recent experimental and theoretical work draws on these results and has shown that the salamander retina may be optimized for prediction \cite{palmer+al_15}.

The efficient representation of predictive information that we see in \cite{palmer+al_15} adds the notion of relevant information to the classical ideas of efficient coding.  The simplest version of the efficient coding hypothesis is that the retina processes visual inputs to remove redundancy, allowing the array of retinal ganglion cells to make fuller use of their limited capacity to transmit information \cite{barlow_61,attneave_54,atick+redlich_92}.  The results in \cite{palmer+al_15} suggest that the retina is not designed to represent all of the input light patterns impinging on its photoreceptors, but instead to represent those parts of the input that are most predictive of the future.  The retina clearly throws away some aspects of the input light patterns, but perhaps only those parts that are irrelevant for the task of prediction.  

\paragraph{Information bottleneck approaches}  The maximal amount of predictive information a system can possibly encode can be found by solving the following information bottleneck problem: 
\begin{equation}
\min_{p(r_t|s_t)} I(R_t;S_{t,t-\Delta t, t-2\Delta t, \ldots}) - \beta I(R_t;S_{t+\Delta t}).
\end{equation}
This can also be understood as a rate distortion problem where the distortion metric is the predictive information.  Here we can see how predictive information and predictive coding are really solving complementary optimization problems for stationary stimuli. 

Providing an efficient representation of predictive information is nearly opposite of what one would expect from neurons doing predictive coding.  In that type of code, signals are decorrelated in time so that  predictable components are eliminated and neurons encode only the deviations from expectation, or  surprise.   The responses of  neurons implementing a truly optimal predictive code in a stationary input environment would thus carry no predictive information about their own responses.  In contrast, recent results \cite{palmer+al_15} suggest that the retina has a large amount of response predictive information, and that these responses efficiently separate predictive from non--predictive bits, and transmit the predictive bits preferentially.

Predictive coding does not explicitly preserve information about the future stimulus.  As such, it is hard to compare to optimal predictive information schemes.  The prescription for predictive coding is if $S$ is the input, and $R^*$ is the expected response and $R$ is the observed response: when $p(r|s)$ is much different from $p(r^*|s)$, respond.  Somewhere in the brain must live the model(s) that generates $r^*$. Higher cortical areas that inhibit early sensory areas, such as the lateral geniculate nucleus, might provide precisely this type of feedback, and have been implicated in experimental evidence for predictive coding, se e.g.\cite{rao+ballard_99}.  In the retina, with little to no feedback from any downstream area, these models must be wired into the retinal circuit.  

\paragraph{Coding for surprising stimuli} 
Predictive coding and optimal predictive information are mostly opposed processing theories.  We illustrate this by way of a few toy examples: If we imagine a world that is wholly static, there is nothing to predict, no predictive information, and a predictive coder would have no response.  If the world is instead completely stochastic, however, predictive coding would dictate that all noise signals that deviate strongly from the prior on the input generate a response, since each one is unexpected and therefore surprising.  These surprising inputs are, however, uninformative about the future, since they are a pure noise signal.  The predictive information present is zero and no response modulation should be encoded.  Predictive coding is, however, designed explicitly for non-stationary stimuli.  Predictive information optimization would code for a surprising change in the inputs, since that will have maximum information about the future state.  In that sense the two methods are aligned and preserving predictive information preferentially over other bits of past information is `efficient'.  

\paragraph{Where models of the input statistics are stored}  Predictive coding, in its clear formulation by Rao and Ballard, postulates that a set of models of the input statistics are present in higher order areas that feedback onto sensory areas to suppress response to predictable events \cite{rao+ballard_99}.  Recent work has elegantly shown how this can be implemented in a hierarchical (Bayesian) framework \cite{deneve_08}.  With limited feedback impinging on the retina, however, the retina itself must store the models of the input statistics it will receive.  It has been demonstrated how adaptive gain mechanisms in the retina, that have presumably been encoded over evolutionary time, can instantiate predictive coding in the retina \cite{hosoya+al_05}.  To further test these ideas in the context of natural scene statistics, one needs to define the set of motion models an organism encounters in its natural environment.  

\section{LNP models fail to capture motion processing in the retina}
In many contexts, our simplest models of retinal ganglion cell firing fail to recapitulate the actual response properties of the retina.  This is most clearly demonstrated for motion stimuli. So called LN, or linear-nonlinear models, of neural processing fail to recapitulate the motion processing properties of the retina for a variety of moving stimuli, including motion anticipation, the reversal response, object motion detection, and the omitted-stimulus response.  Sometimes complex adaptive gain controls mechanisms need to be added to these models to explain the motion processing of the retina. In LN models, the probability of spiking is an instantaneous, nonlinear function of a linearly filtered version of the sensory input.  In the case of retinal ganglion cells that we study here, the inputs are the image or contrast as a function of space and time, $s(\vec{x},t)$. Thus, if we write the probability per unit time of a spike (the firing rate), we have
\begin{equation}
r_{\rm LN}(t) = r_0 g(z),
\end{equation}
where  $r_0$ sets the scale of firing rates, $g(z)$ is a dimensionless nonlinear function, and
\begin{equation}
z(t) =  \int_0^t d\tau \int d^2 x\, f(\vec{x},\tau) s(\vec{x},t-\tau) ;
\end{equation}
where the function $f(\vec{x},\tau)$ is the receptive field.  

\begin{figure}
\includegraphics[width=0.65\linewidth]{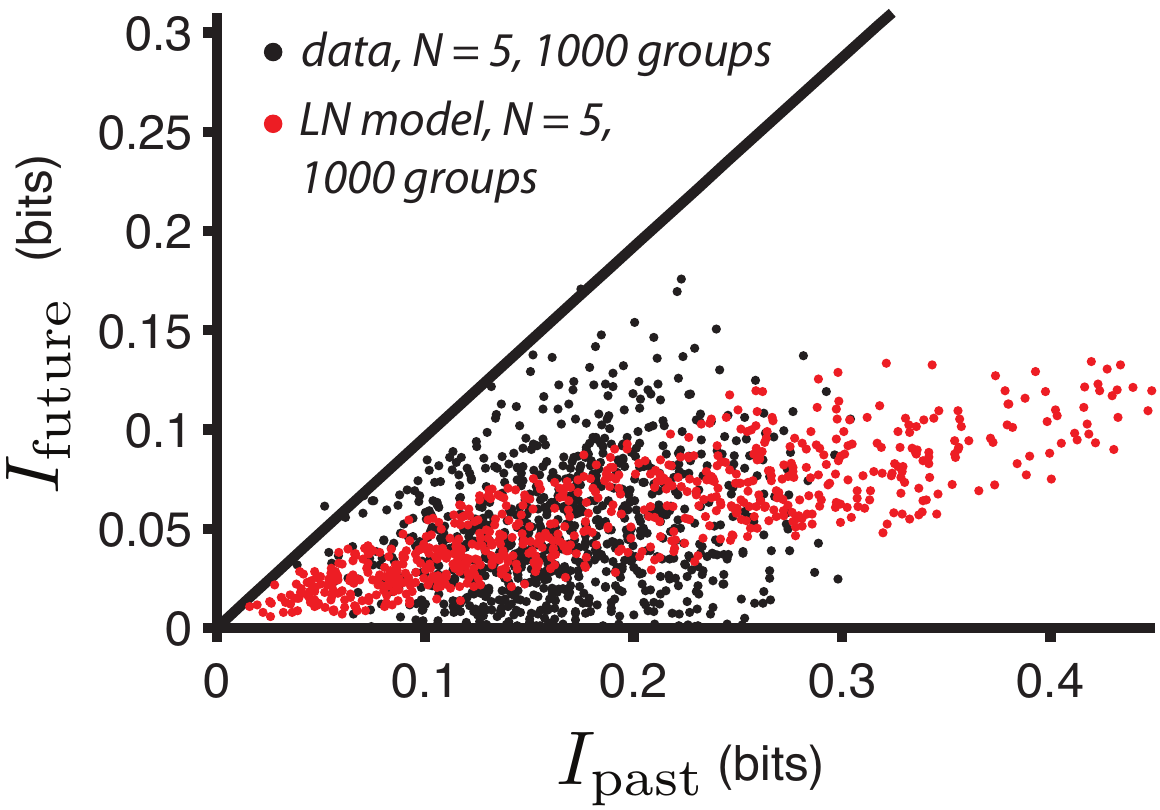}
\caption{The predictive information present in groups of $5$ retinal ganglion cells (RGCs, black dots), as well as for model neurons fit with linear-nonlinear (LN) models (red dots).  The bound on the maximal amount of information about the future a group response with a given about of information about the past stimulus is denoted with the black line.  The input is a moving bar stimulus as in \cite{palmer+al_15}.  Some model groups have more information about the past because the LN model does not capture the low-level noise in the response of the RGC's.  Other model groups have less info than observed in real data because they fail to capture the stimulus driven response of these cells.  No model groups capture as much information about the future of the stimulus as the core predictive groups of the cells in the retina.}
\label{LNmodel}       
\end{figure}

If we deliver stimuli that are drawn from a Gaussian white noise ensemble, then
\begin{equation}
f(\vec{x},\tau) \propto \langle s(\vec{x},t-\tau) \delta(t-t_\mathrm{spike})\rangle ,
\label{revcor}
\end{equation}
where $t_{\rm spike}$ is the time of a spike and $\langle \cdots \rangle$ denotes an average over the stimulus ensemble \cite{chichilnisky_01}. If we take the LN model derived from the random checkerboard stimuli and use it to produce neural responses to the moving bar stimulus, the predictive information carried by the neurons is drastically wrong.  In recent reviews from Gollisch and Meister \cite{gollisch+meister_10} and Berry and Schwartz \cite{berry+schwartz_11}, the myriad ways in which a simple linear-nonlinear-Poisson (LNP) model fails to reproduce known retinal response properties are described in detail.  When simple bars of light move across the retina, reverse their path, blink on and off, or move in more complex ways, many kinds of nonlinear processes in the retina are activated.  None of these effects are captured by this basic version of the LN model for retinal firing. Additionally, results in \cite{palmer+al_15} reveal that the LN model fails to recapitulate the near--optimal behavior of the real data, as illustrated for a slightly different LN model here in Figure \ref{LNmodel}.  

All groups fall away from the bound determined by $\Delta t = 1/60$ s, the delay between the current response and the onset of the common future.  When we compute information about the future, we assume that the future starts now, and do not make any allowances for processing delays.  We could, instead, compare the performance of the LN model with bounds calculated assuming that there is a delay between past and future, so that  $\Delta t^* = \Delta t + t_\textrm{delay}$.  The bound for $\Delta t^*$ is chosen to be $t_\textrm{delay} = 117$ms, comparable to the delay one might estimate from the peak of the information about position, or from the structure of the receptive fields themselves.  Interestingly, the model neurons do come close to this less restrictive bound. 

Of course, these model cells might not be optimal at any delay, instead they could fail to represent all of the predictable components of the stimulus, such as the velocity. Real salamander RGCs have a delay of at least $50$ms, as measured by the time to the peak firing rate induced by a flash.  The data reveal that the retina has a mechanism that allows it to saturate the bound on the predictive information with almost zero effective processing delay when responding to a predictable moving stimulus. 

This work only scratches the surface of optimal coding for the future stimulus in the retina.  The motion statistics chosen here were chosen to include both two time scales of predictable motion as well as a purely stochastic motion component, while still being soluble via Gaussian information bottleneck techniques \cite{chechik+al_05}.  Important extensions of this work will test other parametric motion models with different statistics, but one of the most important directions of future research here is to explore motion models that mimic the properties of natural scenes.

\section{Towards naturalistic motion stimuli}
\begin{figure}
\includegraphics[width=\linewidth]{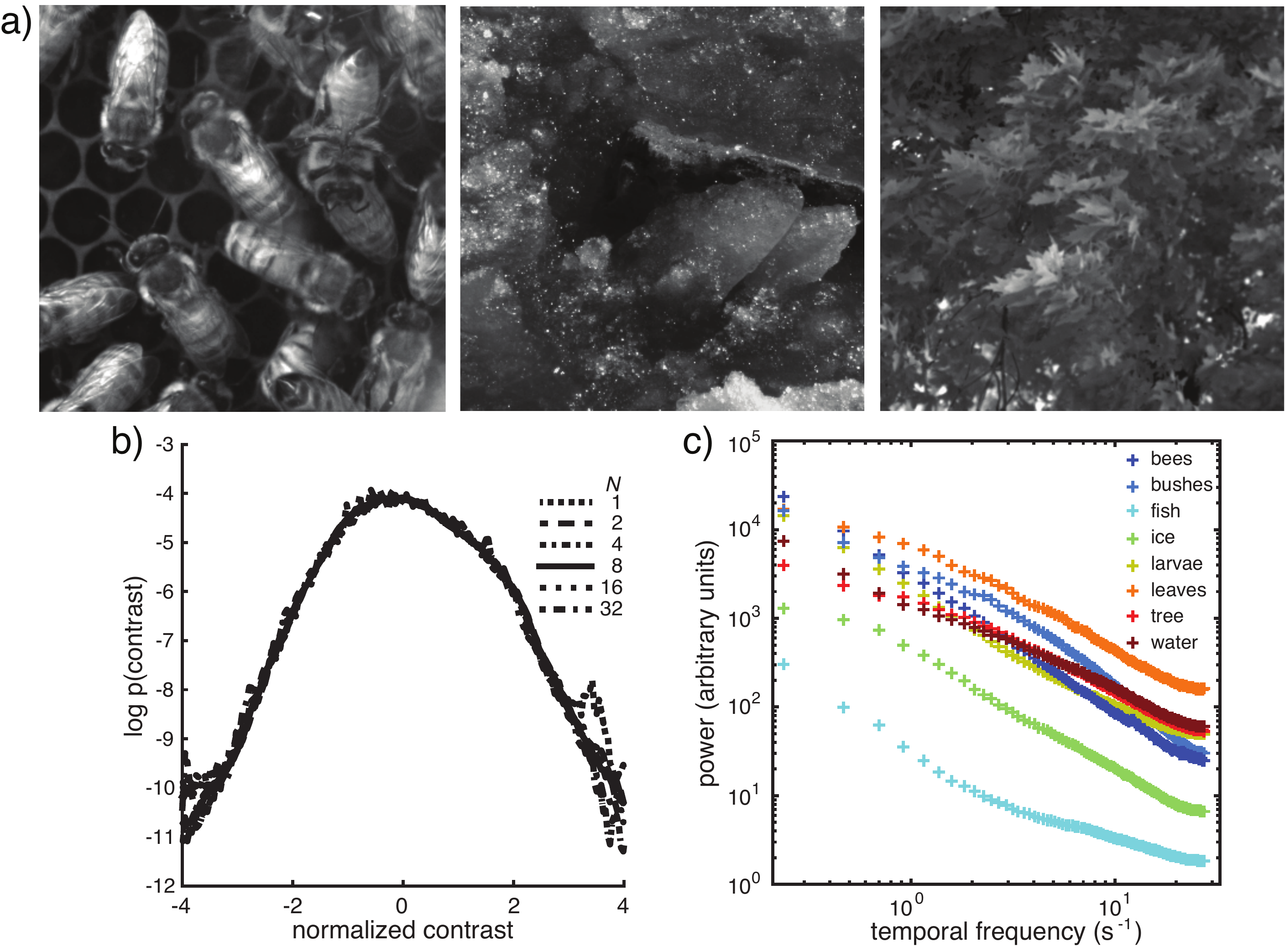}
\caption{(a) Representative frames from 3 natural movies: bees on a honeycomb (a plate of glass exposes the hive from the side); pack ice flowing in Lake Michigan; tree blowing in the wind. (b) Contrast distribution for an ensemble of 8 natural movie clips of 20 s each, filmed at 60 Hz. The contrast of each pixel is defined as $C=I/I_0$ where $I$ is the 8-bit intensity value for that pixel and $I_0$ is chosen for each frame such that the average contrast for that frame is 0. Distributions were estimated at different scales by averaging contrast over $N\times N$ blocks. Contrast distributions are normalized to have unit variance. (c) Temporal power spectra for 8 natural movie clips. Spectra were computed for each pixel using a sliding Hamming window of 256 frames with 50\% overlap, then averaged across pixels.}
\label{natmovies}       
\end{figure}

Natural scenes have heavy-tailed distributions of many quantities of interest, including intensity, contrast, and temporal modulation frequency \cite{olshausen+field_96,dong+atick_95,ruderman_97,billock+al_01}.  This means that there exists no single length scale in space or time one can use to coarse-grain natural scenes without sacrificing large amounts of structure in the data, and that potentially salient fluctuations exist on all scales.  We illustrate some basic statistics of natural scenes in Figure \ref{natmovies}, taken from our own database of natural movies.

To fully test whether the retina conveys information to the brain in a way that is optimized for prediction, we must consider what prediction problems the retina evolved to solve.  In particular, if the retina solves only a subset of the possible spatiotemporal prediction problems that could possibly confront its input, it should solve those that are present in the natural environment.  It could even be the case that the retinas of different animals evolved to encode most efficiently prediction on the scales and correlation structures present in its own ecological niche.  To test this, we need a framework for quantifying the prediction problems present in natural visual stimuli.

\subsection{Quantifying optimal coding for natural motion}
It is not possible to measure the information content of a neuron's response to the pixel-by-pixel representation of an image sequence of any useful size.  Proxies for this calculation include computing information about time within a long and ergodic stimulus sequence \cite{brenner+al_00}, but a better approach is to find some reduced and parametrizable (and therefore sample-able) representation of the motion present in the natural environment that retains the features relevant to the organism.  The heavy-tailed nature of natural scene statistics do not offer up a coarse-graining length scale over which we might smooth our inputs.  Instead, we are left with the difficult task of quantifying and summarizing the key statistics of the natural world.  One useful extraction in the context of prediction is to track salient objects in natural scenes and analyze the statistics of those trajectories.  Another useful experiment would be to show the retina of one animal object trajectories from a different ecological niche and determine whether it fails at this predictive computation while excelling at those drawn from its native environment. 

\paragraph{Higher-order statistics of motion}  Deciding how to quantify a natural scene can be challenging.  The high dimensionality of natural inputs to the visual system means that direct approaches to quantifying the information transmission is wrought with sampling error pitfalls or completely impossible.  Making educated guesses about what features of natural motion to quantify, or searching directly for a lower dimensional representation of the structure of natural scenes are two promising approaches to this problem.  Recent work has defined a set of motion primitives, correlations structures in space and time that find their basis in early theories of motion processing, and extend and complete a local motion structure.  The Fourier, non-Fourier, expander and glider components of local motion can be readily computed from natural movies \cite{hu+victor_10,nitzany+victor_14}.  This approach reveals that certain ratios of these components may be prevalent in natural scenes \cite{nitzany+victor_14}.  The brain might make use of this fact to tailor its motion processing to just these types of input.  If so, deviations from this natural ratio should lead to noisier, less efficient coding for the future stimulus.  Downstream of the retina, this could lead to motion perception deficits.

Such local motion signatures only characterize part of the total motion signal in natural scenes.  Machine learning efforts have been launched to find longer-range, collective components of natural image and motion statistics \cite{cadieu+olshausen_12,hausler+al_12,hausler+al_13,saremi+sejnowski_13}. Work from these groups has shown that some physical models of long range fluctuations may be applicable to natural motion.  This is exciting because it could lead to a generative model for such motion, opening up the possibility of more stringent, parametrized tests of optimal coding for motion in the retina.

\section{Conclusions}
Testing theories of optimal prediction in the visual stream requires an integration of existing theories of optimal coding in the retina and beyond, with a careful quantification of the motion statistics present in the natural environment.  By examining how well information processing in the brain is tuned to natural inputs, we may discover new static and adaptive features of the predictive part of the neural code.

\begin{acknowledgements}
We thank D.\ J.\ Schwab for comments on the manuscript.
\end{acknowledgements}


\bibliographystyle{spmpsci}      

\bibliography{retina_refs_2015}   

\end{document}